\documentclass[conference]{IEEEtran} 
\usepackage{amsmath,graphicx,hyperref}
\usepackage{cite}
\usepackage{subcaption} 
\usepackage{amssymb}
\usepackage{color}
\usepackage{verbatim}
\usepackage{booktabs}
\usepackage{siunitx}
\usepackage{bm}

\renewcommand{\thetable}{\Roman{table}}

\title{NSC-SL: A Bandwidth-Aware Neural Subspace Compression for Communication-Efficient Split Learning}

\author{
    \IEEEauthorblockN{
        Zhen Fang\IEEEauthorrefmark{1},
        Miao Yang\IEEEauthorrefmark{1},
        Zehang Lin\IEEEauthorrefmark{1},
        Zheng Lin\IEEEauthorrefmark{2},
        Zihan Fang\IEEEauthorrefmark{3},  
        Zongyuan Zhang\IEEEauthorrefmark{4}, \\
        Tianyang Duan\IEEEauthorrefmark{4},
        Dong Huang\IEEEauthorrefmark{5} and
        Shunzhi Zhu\IEEEauthorrefmark{1}
    }
    \IEEEauthorblockA{\IEEEauthorrefmark{1}School of Computer and Information Engineering, Xiamen University of Technology, Xiamen, China}
    \IEEEauthorblockA{\IEEEauthorrefmark{2}Department of Electrical and Electronic Engineering, The University of Hong Kong, Hong Kong, China}
    \IEEEauthorblockA{\IEEEauthorrefmark{3}Department of Computer Science, City University of Hong Kong, Hong Kong, China}
    \IEEEauthorblockA{\IEEEauthorrefmark{4}Department of Computer Science, The University of Hong Kong, Hong Kong, China}
    \IEEEauthorblockA{\IEEEauthorrefmark{5}School of Computing, National University of Singapore, Singapore}
}

\begin{document}

\maketitle

\begin{abstract}
The expanding scale of neural networks poses a major challenge for distributed machine learning, particularly under limited communication resources. 
While split learning (SL) alleviates client computational burden by distributing model layers between clients and server, it incurs substantial communication overhead from frequent transmission of intermediate activations and gradients. 
To tackle this issue, we propose NSC-SL, a bandwidth-aware adaptive compression algorithm for communication-efficient SL. NSC-SL first dynamically determines the optimal rank of low-rank approximation based
on singular value distribution for adapting real-time bandwidth constraints. 
Then, NSC-SL performs error-compensated tensor factorization using alternating orthogonal iteration with residual feedback, effectively minimizing truncation loss. 
The collaborative mechanisms enable NSC-SL to achieve high compression ratios while preserving semantic-rich information essential for convergence. Extensive experiments demonstrate the superb performance of NSC-SL.
\end{abstract}


\section{INTRODUCTION}
\label{sec:intro}

With the rapid application of data-driven machine learning in healthcare, finance, and mobile internet~\cite{peng2024sums,lin2024split,yuan2025constructing,zhang2025state,lin2022channel,duan2025llm,fang2024ic3m,lin2024splitlora,yuan2024satsense,zhang2024fedac}, the traditional centralized training paradigm is gradually questioned due to privacy leakage and compliance risks caused by raw data aggregation \cite{kairouz2021advances,hu2024accelerating}.
Federated learning (FL) \cite{FL,fang2024automated} thus emerges as a distributed privacy-preserving alternative, allowing clients to train models locally and share only parameter updates with a server.
While FL mitigates issues such as data silos and compliance, it faces communication challenges: large-scale models require exchanging substantial parameters each round, leading to high overhead under limited bandwidth and latency \cite{Yang2021}.

To mitigate these issues, split learning (SL) \cite{lin2025esl,wei2025pipelining,lyu2023optimal} emerges as a promising alternative.
SL partitions a deep neural network between clients and a server at a specific layer. Clients perform forward propagation only up to the cut layer and transmit the intermediate activations (known as “smashed data”) to the server, which executes the remaining forward and backward passes and returns the corresponding gradients.
This significantly reduces computational and memory requirements on edge devices, making SL particularly suitable for resource-constrained environments or large-scale models \cite{lin2023pushing,SLFHI}.
However, SL introduces new challenges: it requires frequent per-mini-batch client-server communication during both forward and backward propagation, rendering it highly sensitive to network latency and instability \cite{MuShen2023}.

To mitigate communication bottlenecks in SL, various compression strategies have been proposed~\cite{LPLR,PowerSGD,QSGD,RandTop-k}.
Low-rank approximation techniques, such as ACPSGD~\cite{ACPSGD} and PowerSGD~\cite{PowerSGD}, compress data via low-dimensional projection but typically rely on fixed ranks, which fail to adapt to dynamic network conditions.
Quantization approaches, such as QSGD~\cite{QSGD}, reduce representation precision but often incur accumulated quantization errors and require careful parameter tuning.
Similarly, sparsification techniques like RandTopk~\cite{RandTop-k} transmit only elements with large magnitudes; however, they are prone to information loss and training instability if the sparsity level is not appropriately configured.
Crucially, these existing methods generally do not explicitly account for real-time bandwidth constraints or dynamically adjust compression strategies, thereby limiting their applicability in practical distributed systems.

To address the above issue, in this paper, we propose a bandwidth-aware \underline{n}eural \underline{s}ubspace \underline{c}ompression-based \underline{SL}, named NSC-SL, to substantially reduce the transmission overhead of smashed data while preserving training stability. NSC-SL comprises two key components: bandwidth-aware adaptive rank selection (BAS) and orthogonal alternating subspace approximation (OASA) with error correction loop (ECL). BAS dynamically determines a near-optimal compression rank based on real-time bandwidth constraints and singular value estimation, while OASA iteratively enhances low-rank approximation through residual feedback to minimize truncation errors.
The main contributions of this paper are summarized as follows:
\begin{itemize}
\item We develop BAS to dynamically determine the compression rank by jointly optimizing the spectral energy and real-time communication constraints, which preserves critical information for SL.
\item We design OASA with ECL to achieve efficient and stable low-rank approximation without full singular value decomposition (SVD), significantly reducing computational complexity and error accumulation.
\item We empirically evaluate NSC-SL with extensive experiments, demonstrating that the proposed NSC-SL outperforms state-of-the-art benchmarks.
\end{itemize}
The rest of the paper is organized as follows. Sec.\ref{sec:system design} presents the system design of NSC-SL. Sec.\ref{sec:performance evaluation} presents the performance evaluation, highlighting the superiority of NSC-SL.
Finally, conclusions are presented in Sec.\ref{sec:conclusion}.

\section{SYSTEM DESIGN}
\label{sec:system design}

\subsection{Overview}

\begin{figure}[t]
  \centering
  \includegraphics[width=\linewidth, keepaspectratio]{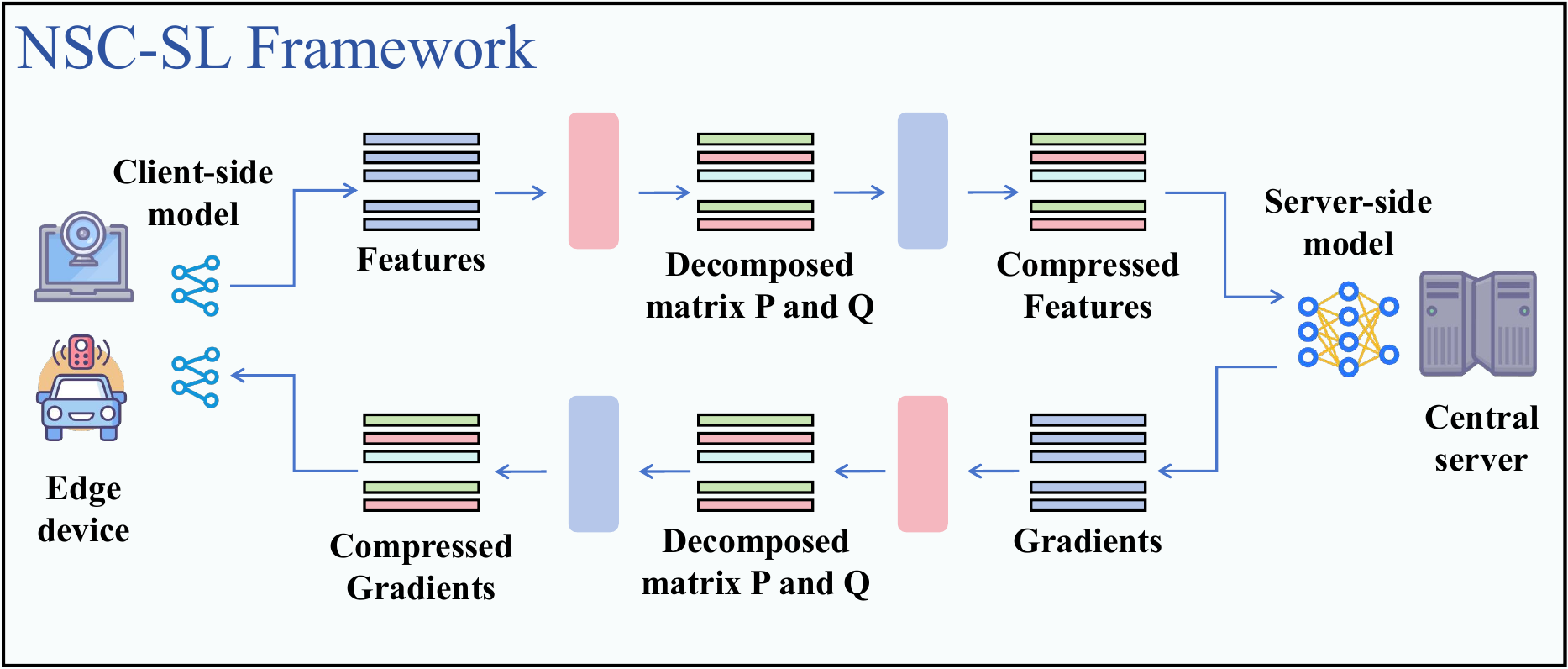}
  \vspace{-1em}
  \caption{Architecture of the proposed NSC-SL framework.}
  \label{fig:nsc-sl-framework}
  \vspace{-1em}
\end{figure}

In this section, we introduce the architecture and workflow of the proposed NSC-SL framework.
In SL, the global model is divided into a client-side submodel and a server-side submodel, which are deployed on clients and edge servers, respectively.
As shown in Fig.~\ref{fig:nsc-sl-framework}, the system consists of two basic parts: i) \textbf{clients}: The set of participating clients is denoted as $\mathcal{N}=\{1, 2, ..., N\}$. The $n$-th client holds a local submodel with parameters $W_{e}^{n}$ and a local dataset $D_n$; ii) \textbf{Server}: The edge server is a computationally powerful entity, with server-side submodel parameters denoted $W_s$. During training, the server receives smashed data from clients and transmits the gradients to clients.

The training workflow of NSC-SL in each training round comprises the following four stages: i) Each client performs forward propagation of its client-side submodel to generate the activations; ii) The generated activations are compressed by the NSC module via low-rank approximation and transmitted to the edge server.
iii) The edge server performs decompression to recover the approximate activations and completes the remaining forward and backward propagation of the server-side submodel to obtain gradients. Subsequently, the gradients are compressed again on the server side to reduce the overhead of downlink transmission. iv) The client receives the decompressed gradients and uses them to update the parameters of the local submodel. Meanwhile, residual feedback is updated locally to provide more accurate corrections for compression in the next training round.

\vspace{-0.6em}
\subsection{Adaptive Rank Selection}
\vspace{-0.6em}
Most existing low-rank compression methods adopt a fixed or empirically chosen rank \cite{halko2011finding, PowerSGD,lin2025sl}, which poses significant limitations in practical training.
On the one hand, tensors from different network layers and batches vary greatly in dimension, energy distribution, and redundancy; thus, a fixed rank often results in either resource waste or accuracy loss.
On the other hand, these methods usually ignore the bandwidth constraints inherent in distributed systems, where the compressed communication volume may exceed the network budget, thereby affecting the feasibility of training \cite{LinHM0D18}.
To overcome these limitations, we introduce an adaptive rank selection mechanism that explicitly considers both energy coverage and bandwidth overhead, as well as the rank cap.
This ensures that the selected rank not only dynamically captures the dominant information of the current tensor, but also satisfies bandwidth constraints.

To determine the compression rank, we first approximate the singular value spectrum of the target matrix $\bm{M}\in\mathbb{R}^{m\times n}$.
While a full SVD provides an exact solution, its computational complexity of $O(\min(m,n)^3)$ is often prohibitive for large-scale matrices. Instead, we adopt a randomized subspace method for efficient spectral estimation \cite{halko2011finding}.
We begin by generating a Gaussian random matrix ${\Omega}\sim\mathcal{N}(0,1)^{n\times (r+p)}$ where $p$ is an oversampling parameter. We then perform an iterative orthogonalization process:
\begin{equation}
\bm{Y}_q = 
\begin{cases}
  \bm{M}\bm{\Omega}, & q = 0, \\
  (\bm{M}\bm{M}^\top)^q \bm{Y}_0, & q \ge 0.
\end{cases}
\end{equation}

Next, we compute a thin QR decomposition $\bm{Y}_q=\bm{Q}\bm{R}$ with $\bm{Q}^\top \bm{Q}=\bm{I}$.
We then project and factorize to approximate the spectrum and subspaces.
\begin{equation}\label{eq:factor}
\begin{aligned}
\bm{B} &= \bm{Q}^\top \bm{M} \in \mathbb{R}^{(r+p)\times n}, \bm{B} = \tilde{\bm{U}}\bm{\Sigma}\bm{V}^\top.
\end{aligned}
\end{equation}

The spectrum is approximated by $\sigma_i(\bm{M}) \approx \bm{\Sigma}_{ii}$,
and the dominant singular subspaces are approximated by
$\bm{U}_r \approx \bm{Q}\,\tilde{\bm{U}}_{(:,1:r)}$ 
and $\bm{V}_r \approx \bm{V}_{(:,1:r)}$.
The computational complexity is reduced to $O(mnr)$,
where $r \ll \min{m,n}$, providing a near-SVD spectral estimate
with significantly lower complexity.

Let the singular values of matrix $\bm{M}$ be ordered descendingly as
$\sigma_1(\bm{M}) \geq \sigma_2(\bm{M}) \geq \cdots \geq \sigma_{\min(m,n)}(\bm{M})$.
To balance information preservation and communication efficiency during compression,
we use the \textit{energy coverage} ratio as a criterion.
Specifically, for a given threshold $\eta \in (0,1)$, the optimal compression rank $r_\eta$
is defined as the smallest integer $r$ such that the cumulative energy reaches at least $\eta$ fraction of the total energy:
\begin{equation}\label{eq:adaptive_rank1}
r_{\eta} \;=\; \min \left\{ r \;\middle|\;
\frac{\sum_{i=1}^{r} \sigma_i^2(\bm{M})}{\sum_{i=1}^{\min(m,n)} \sigma_i^2(\bm{M})} \;\ge\; \eta \right\}.
\end{equation}

This criterion ensures that the low-rank approximation $\hat{\bm{M}}$ 
retains at least $\eta$ portion of the original energy.

To integrate practical system constraints into the rank selection process, we incorporate both a communication budget and a predefined rank cap. The communication-aware rank is derived from the maximum allowable transmission size $B_{\text{max}}$ (in bytes), accounting for the fact that transmitting two factor matrices $\bm{P}$ and $\bm{Q}$ in single-precision (32-bit, i.e., 4 bytes per value) requires \( 4r(m + n) \) bytes. To ensure the transmitted data size does not exceed \( B_{\text{max}} \), the rank must satisfy:
\begin{align}
r \leq \left\lfloor \frac{B_{\text{max}}}{4(m + n)} \right\rfloor.
\end{align}

Additionally, a predefined rank cap \( r_{\text{cap}} \) is imposed to prevent excessive computational load or numerical instability. Thus, the final compression rank \( r \) is determined by:
\begin{equation}
r = \min \left\{ r_{\eta}, \left\lfloor \frac{B_{\text{max}}}{4(m + n)} \right\rfloor, r_{\text{cap}} \right\}.
\label{eq:adaptive_rank2}
\end{equation}

This integrated selection mechanism simultaneously adapts to the real-time communication constraints, which achieves a tunable trade-off between approximation accuracy and communication efficiency.

\subsection{Orthogonal Alternating Subspace Approximation with Error Correction Loop}
Although truncated SVD theoretically provides the optimal approximation, its high computational complexity and poor scalability make it unsuitable for repeated use in large-scale distributed training \cite{li2019svdcomplexity}.
Alternating orthogonal iteration updates the left and right factor matrices alternately while maintaining their orthogonality, yielding a stable low-rank subspace approximation with lower complexity.
When the target rank is much smaller than the matrix dimension, this approach can significantly reduce computational cost while avoiding the numerical instability caused by one-shot truncation.
However, any low-rank approximation inevitably discards part of the information, and if left unaddressed, such errors may accumulate over multiple rounds of communication and iteration, thereby degrading model convergence performance \cite{SeideFDLY14}.
To address these limitations, we introduce an enhanced strategy consisting of two parts: OASA for efficient low-rank factorization, and ECL to iteratively reduce approximation error.

Given a matrix $\bm{M}$ and a target rank $r$,
the algorithm begins by constructing an initial orthonormal basis $Q^{(0)}$ via random initialization followed by orthogonalization. The method then proceeds iteratively, alternately updating the left and right subspace bases according to the following rules:
\begin{equation}\label{eq:oasa1}
\begin{aligned}
\bm{P}^{(t)} &\;\leftarrow\; \phi\ \!\big( \,(\bm{M}+\bm{E}^{(t-1)})\, \bm{Q}^{(t-1)} \,\big), \\
\bm{Q}^{(t)} &\;\leftarrow\; \phi\ \!\big( \,(\bm{M}+\bm{E}^{(t-1)})^{\top} \bm{P}^{(t)} \,\big).
\end{aligned}
\end{equation}

$\phi$ denotes the orthogonalization operation, where $\mathbf{E}^{(t)}$ denotes the residual feedback term of error correction at iteration $t$. Therefore, the low-rank approximation at the $t$-th iteration is given by:
\begin{equation}
\hat{\bm{M}}^{(t)} \;=\; \bm{P}^{(t)} \big(\bm{Q}^{(t)}\big)^{\top}.
\label{eq:oasa2}
\end{equation}
Then, we update the residual feedback as:
\begin{equation}
\bm{E}^{(t)} \leftarrow \beta \bm{E}^{(t-1)} + \big(\bm{M} - \hat{\bm{M}}^{(t)}\big),
\label{eq:oasa3}
\end{equation}
where $\beta \in [0, 1]$ is a momentum coefficient that controls the reinjection of historical truncation errors, thereby enhancing stability and convergence.
Compared to conventional truncated SVD with a complexity of $O(\min(m,n)^3)$, the per-iteration complexity of the proposed algorithm is:
\begin{equation}
O(mnr + (m+n)r^2) \;\approx\; O(mnr),
\label{eq:oasa4}
\end{equation}
which significantly reduces the computational cost as $r \ll \min(m,n)$.

After $T$ iterations, the final approximation is constructed as:
\begin{equation}
\hat{\bm{M}} = \bm{P}^{(T)} \big(\bm{Q}^{(T)}\big)^{\top},
\label{eq:approximate_matrix}
\end{equation}
which converges stably toward the optimal rank-$r$ approximation in the Frobenius norm:
\begin{equation}
\hat{\bm{M}} \;\approx\; \arg\min_{\operatorname{rank}(\bm{X})=r} \;\|\bm{M} - \bm{X}\|_F,
\label{eq:Optimal_approximation}
\end{equation}

To enhance computational efficiency, we introduce an early-stopping criterion based on relative residuals. After each iteration, the relative residual is evaluated. If its improvement over the best recorded value is below a predefined threshold, a stagnation counter is incremented; otherwise, the best residual is updated, and the counter is reset. Training stops early when the counter exceeds a patience parameter after a minimum number of iterations. This approach reduces redundant computation while preserving approximation quality.

\vspace{-0.9em}
\section{PERFORMANCE EVALUATION}
\label{sec:performance evaluation}
\vspace{-0.6em}
In this section, we introduce the experimental setup and evaluate NSC-SL against state-of-the-art benchmarks.
\vspace{-0.8em}
\subsection {Implementation Settings}
\vspace{-0.6em}

\textbf {Dataset and Model:}  
The performance of NSC-SL is evaluated on the HAM10000 \cite{DVN/DBW86T_2018} skin lesion image classification dataset under IID data partitioning. For model deployment, ResNet-18 \cite{ResNets} is adopted as the global network and is partitioned according to the SL architecture: the first 3 layers are deployed on clients as the client submodel, and the remaining layers are deployed on the server as the server submodel.

\textbf{Baselines:}
To comprehensively evaluate the performance of NSC-SL, we compare it with the following baselines:
i) ACPSGD \cite{ACPSGD}, which alternates the low-rank compression and aggregation between PowerSGD's P and Q;
ii) RandTopk \cite{RandTop-k}, a top-k SL variant that compresses smashed data by retaining the top-k elements with the highest magnitude and a small portion of randomly selected non-top-k elements;
iii) QSGD \cite{QSGD}, an efficient quantization approach that reduces the accumulating quantization errors.

\textbf{Hyperparameters:}
In our experiments, we construct an experimental scenario with 5 clients and one edge server.
We use the stochastic gradient descent (SGD) optimizer with a learning rate set to 0.0001.
The mini-batch size is set to 128.
The system bandwidth is set to 100 Mbps, and transmission latency to 50 ms.

\subsection{Superiority of NSC-SL}

\begin{figure}[t]
  \centering
  \begin{subfigure}[t]{0.45\linewidth}
    \centering
    \includegraphics[width=\linewidth, keepaspectratio]{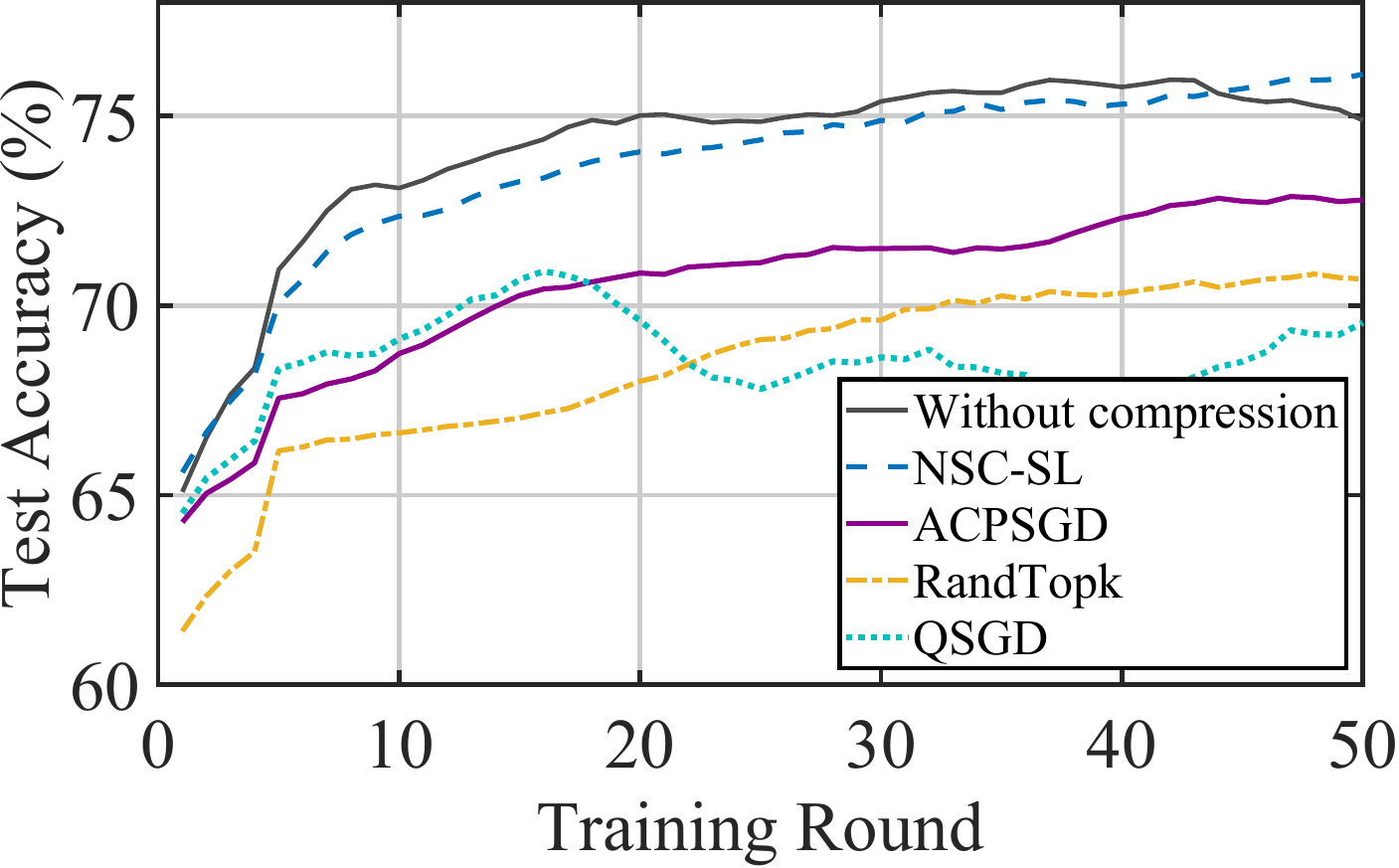}
    \caption{Test accuracy over rounds.}
    \label{fig:acc_test}
  \end{subfigure}
  \hfill
  \begin{subfigure}[t]{0.455\linewidth}
    \centering
    \includegraphics[width=\linewidth, keepaspectratio]{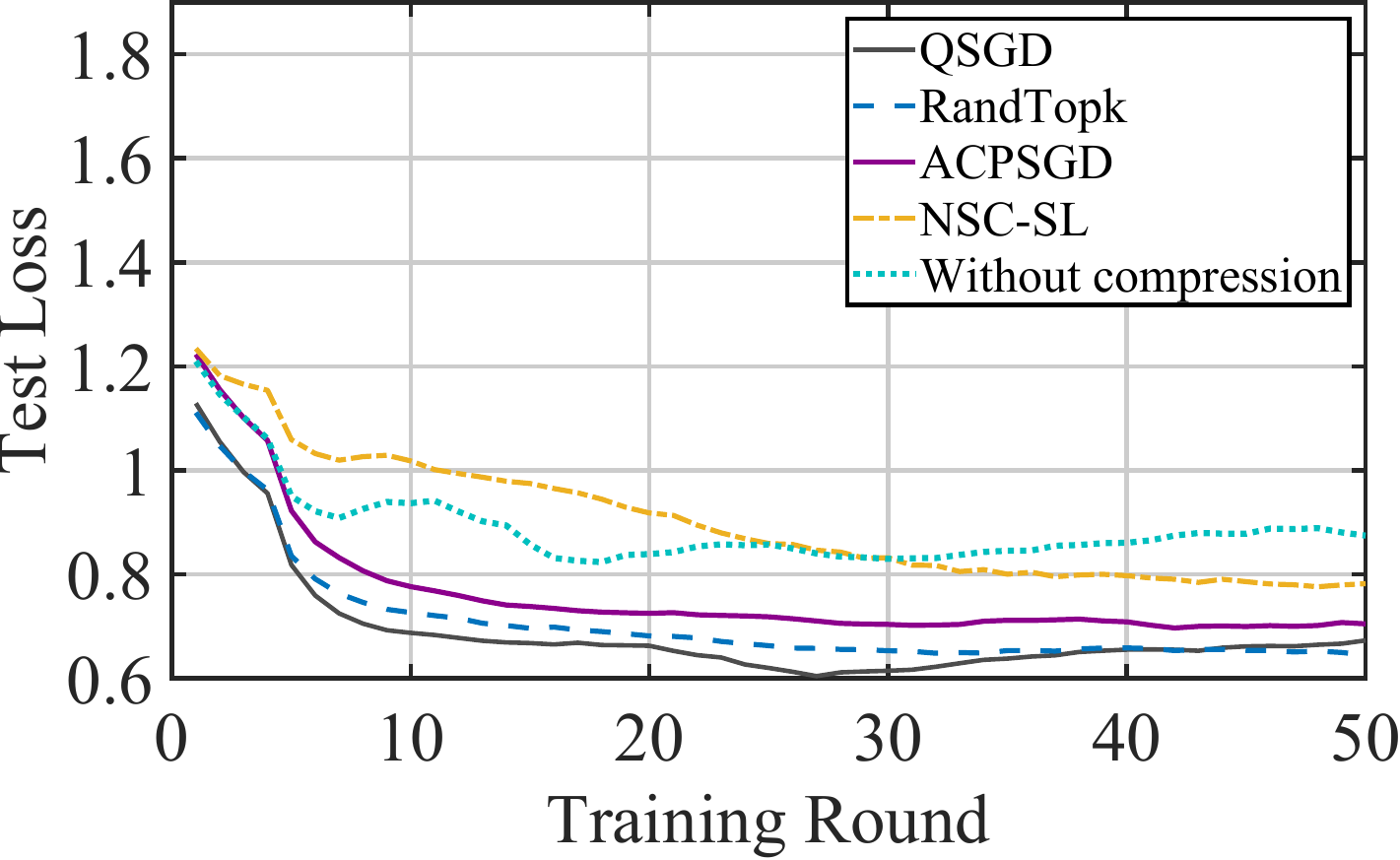}
    \caption{Test loss over rounds.}
    \label{fig:loss_test}
  \end{subfigure}
  \vspace{-0.6em}
  \caption{The test performance of ResNet-18 on the HAM10000 datasets.}
  \label{fig:test_performance}
  \vspace{-1em}
\end{figure}

\begin{table}[t]
\centering
\caption{The MSE accuracy of the compression with bandwidth limitation on the HAM10000 dataset.}
\scalebox{0.8}{ 
\begin{tabular}{l *{4}{S[table-format=1]}}
\toprule
Algorithm & \multicolumn{1}{c}{25\,Mbps} & \multicolumn{1}{c}{50\,Mbps}
          & \multicolumn{1}{c}{100\,Mbps} & \multicolumn{1}{c}{200\,Mbps} \\
\midrule
RandTopK & 0.68 & 0.56 & 0.43 & 0.28 \\
ACPSGD   & 0.24 & 0.21 & 0.18 & 0.14 \\
QSGD     & 1.44 & 0.66 & 0.19 & \bfseries0.06 \\
NSC-SL   & \bfseries 0.17 & \bfseries 0.14 & \bfseries 0.11 & \bfseries0.06 \\
\bottomrule
\end{tabular}
}
\label{tab:mse_results}
\end{table}

Fig.~\ref{fig:acc_test} and Fig.~\ref{fig:loss_test} present the test accuracy and test loss of baseline algorithms on the HAM10000 dataset.
The proposed NSC-SL framework consistently outperforms all baseline methods, demonstrating its superb performance. This performance advantage stems from NSC-SL’s enhanced computational and system efficiency: the integration of BAS and OASA promotes stable and rapid convergence, while power iterations and adaptive oversampling improve spectral estimation accuracy.
In contrast, methods such as RandTopk exhibit a noticeable performance gap, primarily because of their limited representational capacity and discarding of global structural correlations. Similarly, ACPSGD and QSGD also fall short of NSC-SL, as they fail to adequately preserve long-range dependencies and underlying data manifolds in the compressed representations.

Table.~\ref{tab:mse_results} reports mean squared error (MSE) and accuracy under varying bandwidths. For fair comparison, NSC-SL is evaluated without error feedback. NSC-SL consistently outperforms baselines, reducing MSE by 18.3\%–42.7\% and improving accuracy by 4.1\%–11.5\%. The gain comes from randomized subspace estimation, which adaptively preserves dominant spectral components under aggressive rank reduction, whereas quantization and sparsification discard information indiscriminately, leading to distortion and accuracy loss.

\subsection{Ablation Experiments}

\begin{figure}[t]
  \centering
  \begin{subfigure}[t]{0.45\linewidth}
    \centering
    \includegraphics[width=\linewidth, keepaspectratio]{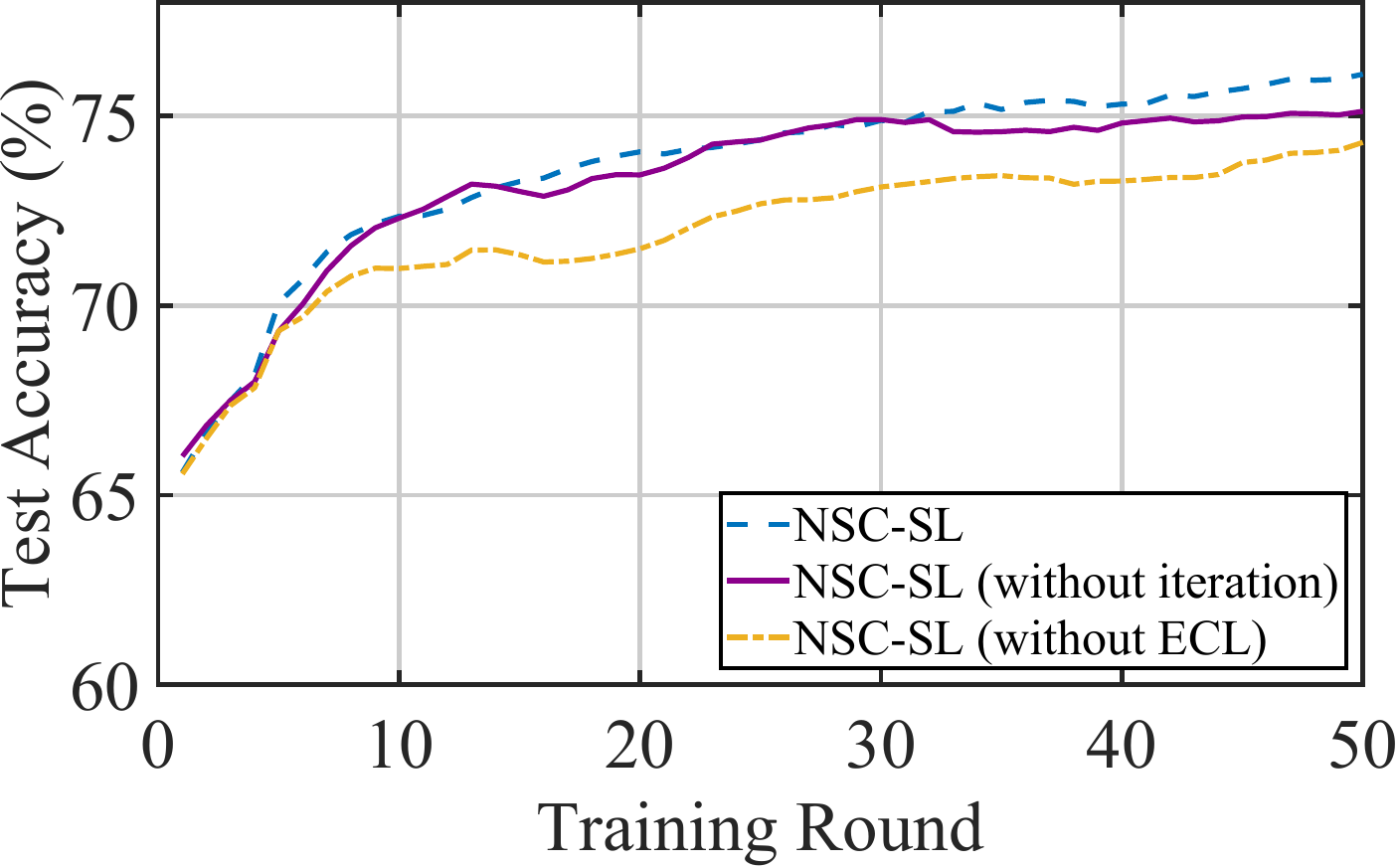}
    \caption{Test accuracy.}
    \label{fig:acc_test_melt}
  \end{subfigure}
  \hfill
  \begin{subfigure}[t]{0.44\linewidth}
    \centering
    \includegraphics[width=\linewidth, keepaspectratio]{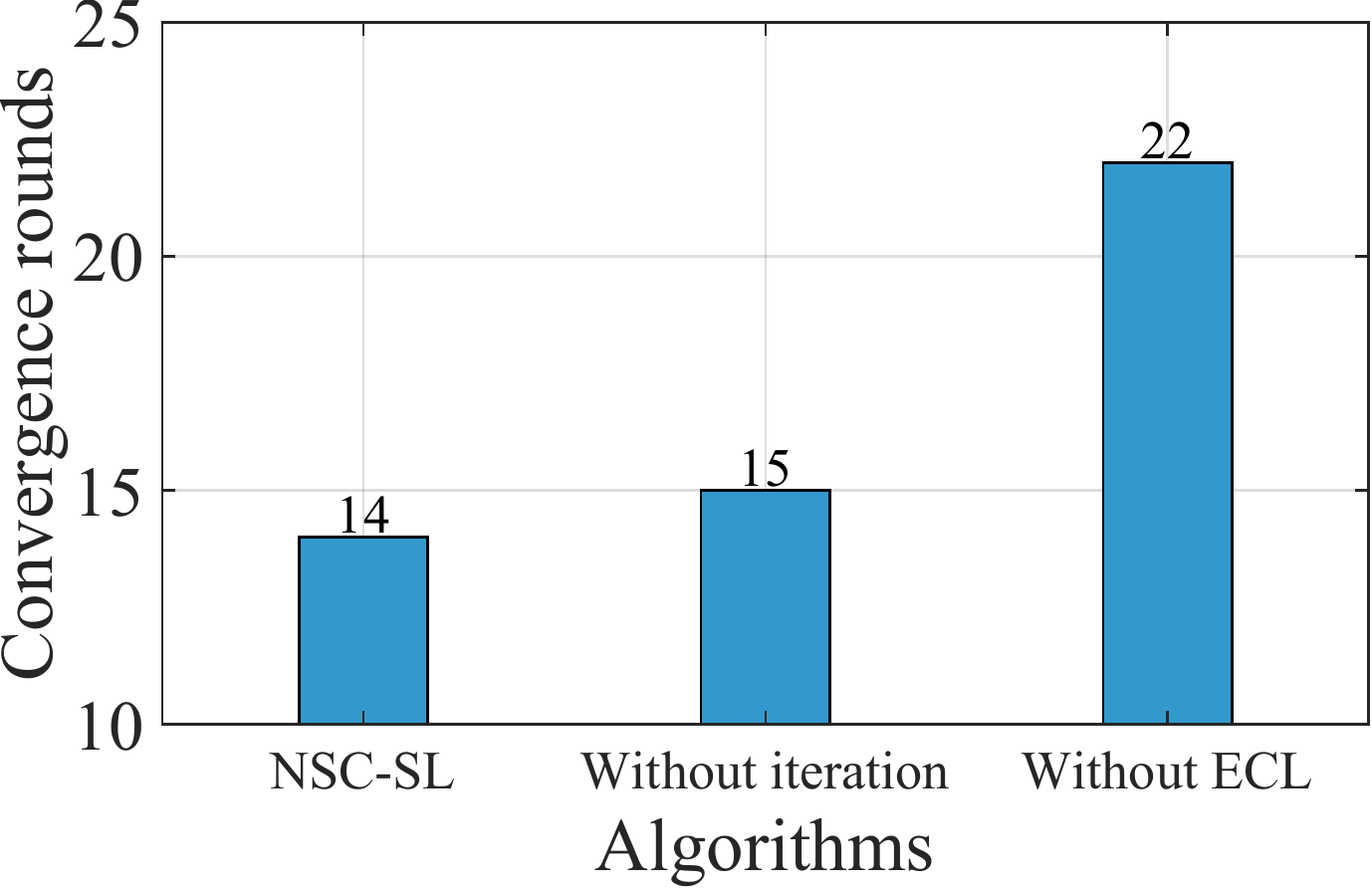}
    \caption{Model convergence.}
    \label{fig:number_of_convergence_rounds}
  \end{subfigure}
  \vspace{-0.6em}
  \caption{The ablation experiments for OASA on the HAM10000 dataset without iteration and without ECL.}
  \vspace{-0.6em}
  \label{fig:combined}
\end{figure}

We conduct an ablation study to evaluate the core components of OASA. As shown in Fig.~\ref{fig:acc_test_melt} and Fig.~\ref{fig:number_of_convergence_rounds}, removing ECL results in an accuracy drop of 1.72\%. This is primarily because ECL utilizes residual errors to guide subspace updates and suppresses systematic bias. Without ECL, residuals accumulate over iterations, increasing reconstruction error and gradient bias, which in turn reduces optimization progress and triggers early stopping. Under bandwidth or energy constraints, the effective rank is further reduced, worsening underfitting.

Similarly, removing iteration leads to a 0.91\% decrease in accuracy. Without the multi-step stabilization provided by OASA, a single alternation between factors
$\bm{P}$ and $\bm{Q}$ are insufficient to achieve a stable subspace, resulting in higher reconstruction error. Under rank truncation, more iterations are required for convergence; without them, both underfitting and bias are amplified.

\section{CONCLUSION}

\label{sec:conclusion}

This paper presents NSC-SL, a communication-efficient SL framework that significantly reduces communication overhead while preserving model performance. NSC-SL comprises two key components: BAS, which dynamically adjusts compression levels based on real-time bandwidth constraints, and OASA, which mitigates truncation loss via iterative refinement with residual feedback. The collaborative components of NSC-SL ensure high compression ratios without sacrificing critical semantic information required for convergence.

\bibliographystyle{IEEEbib}
\bibliography{refs}
\label{sec:references}
\end{document}